\documentclass[a4paper,preprint,5p,number,sort,compress]{elsarticle}

\usepackage{amsmath}
\usepackage{amssymb}
\usepackage{graphicx}
\usepackage{dcolumn}
\usepackage{bm}
\usepackage[hidelinks]{hyperref}

\hypersetup{
  colorlinks   = true,
  urlcolor   = blue,
  citecolor   = blue
}

\begin{document}

\title{Optical, vibrational, and electronic properties of semiconducting YbN}

\author[1]{M.~Markwitz\corref{cor1}}\ead{martin.markwitz@vuw.ac.nz}
\author[1]{C.~Pot}
\author[1,2]{R.~G.~Buckley}
\author[1,2]{W.~F.~Holmes-Hewett}
\author[1,2]{S.~Granville}

\affiliation[1]{Robinson Research Institute, Victoria University of Wellington, P.O. Box 33436, Petone 5046, New Zealand}
\affiliation[2]{The MacDiarmid Institute for Advanced Materials and Nanotechnology, Victoria University of Wellington, P.O. Box 600, Wellington 6140, New Zealand}

\cortext[cor1]{Corresponding author}

\begin{abstract}
    We investigate the vibrational, optical, and electronic properties of insulating YbN thin films using Raman spectroscopy, Fourier-transform infrared spectroscopy, and electrical transport measurements, supported by density functional theory. Raman spectra reveal the LO($\Gamma$) phonon and a cation‑vacancy mode, while the optical conductivity identifies the TO phonon and an absorption edge corresponding to a $1.7$~eV N~2$p\rightarrow$Yb~5$d$ transition. The films exhibit thermally activated resistivity consistent with an insulating ground state. An additional defect induced absorption tail below the intrinsic band gap is observed, which in combination with the electrical measurements indicates the Fermi energy resides in a disordered conduction band minimum.
\end{abstract}

\begin{keyword}
    Rare earth nitride \sep optical conductivity \sep semiconductor \sep electronic structure.
\end{keyword}

\begin{highlights}
\item Raman spectroscopy reveals LO phonon and ytterbium vacancy‑induced local mode
\item Optical conductivity identifies TO phonon and absorption edge
\item Transport measurements show insulating behavior with activation energy-type resistivity 
\item Antiferromagnetic band structure calculation shows indirect 5$d\rightarrow$2$p$ band gap and 4$f$ states just above conduction band minimum
\end{highlights}

\maketitle


The rare earth nitrides (\textit{Ln}Ns) are a series of rock-salt semiconductors which have been subjected to intense investigation for their potential in a wide array of applications~\cite{Natali2013,holmes2025rare}. They exhibit a broad range of magnetic behaviors at cryogenic temperatures based on the choice of \textit{Ln} due to the difference in localized 4$f$ state occupation as the series is traversed. The clearest example of one finding an application is GdN, due to its combination of resistivity and large spin angular momentum which is of interest in the field of superconducting spintronics research~\cite{birge2024ferromagnetic,Caruso2019,zhu2017superconducting,Pot2023,vilela2024spin,pal2014pure,massarotti2015macroscopic,ahmad2020electrodynamics}. ErN, on the other hand has found potential application in cryocooling based on its large specific heat at liquid helium temperatures~\cite{nakagawa2017optimization,ye2026analysis}. 

It is only in the past two decades that almost all members of the \textit{Ln}Ns have been found to exhibit properties that resemble an insulating rather than metallic ground states. This is due to historical confusion originating experimentally from contamination and unintentional doping during sample preparation. Advances in preparation conditions, such as powder preparation and characterization entirely within a glove box, or thin film preparation in an ultra high vacuum deposition system ($\leq10^{-9}$~mbar) with an appropriately thick capping layer have been used to study the \textit{Ln}Ns with low levels of contamination~\cite{Kneisel2024,trewick2025conductivity,chen2025epitaxial,loyal2023coexistence}. Often, theory can provide insight into the ground state of a compound, however in the \textit{Ln}Ns an inappropriate description of the exchange and correlation by use of just the local density or generalized gradient approximations in density function theory (DFT) generally predicted metallic ground states. Employment of self interaction-corrected, Hubbard $U$, quasiparticle self-consistent GW, or (screened) exact-exchange or approximations thereto provide a description that is generally found to be consistent with experimental results of the \textit{Ln}N ground states as insulators~\cite{aerts2004half,larson2006electronic,larson2007electronic,richter2011electronic,markwitz2026optolectronic,markwitz2025raman,galler2022electronic}.

Ytterbium nitride (YbN), the smallest of the magnetic \textit{Ln}Ns with a lattice constant of $4.78$\,{\AA} is the only member with an antiferromagnetic (AFM) ground state~\cite{warring2014ybn,degiorgi1990electronic,donni1990fcc}. A further study found that YbN obeys a type-III rock salt antiferromagnetic ordering with long-range ordering along the $\vec{k}=[1,0,0.5]$ crystal axis when below the Néel temperature $T_{\text{N}}=0.790\pm0.005$~K, similar to YbAs~\cite{donni1989long,donni1990fcc}. Moving now to the vibrational and structural properties, Degiorgi~\textit{et~al.}~\cite{degiorgi1990electronic} report optical measurements finding a TO($\Gamma$) mode frequency of $275$~cm\textsuperscript{-1}, a plasma frequency of $0.2$~eV, and an enhanced optical absorption above $0.8$~eV. A recent investigation using Fourier transform infrared spectrometry (FTIR) by Holmes-Hewett~\textit{et~al.}~\cite{holmes2022gamma} on a broad range of \textit{Ln}N thin films grown by MBE found a similar TO($\Gamma$) mode frequency in thin-film YbN. The recent studies on thin film YbN found a variation in lattice constant ($4.78-4.82$~{\AA}), band gap ($1.5-1.7$~eV), mixed evidence for a multivalent Yb\textsuperscript{2+/3+} state, and plasma frequency, suggesting that the ground state properties of YbN are not yet settled~\cite{warring2014ybn,loyal2023coexistence,chen2025epitaxial}. 

In this work we study and compare the properties of low-conductivity YbN thin films to previous reports and compare YbN to the other \textit{Ln}Ns in the series. We investigate its structural, vibrational, and (opto)electronic properties. From optical measurements and DFT computation with Hubbard energy correction, we find that our low-conductivity YbN has a direct 2$p\rightarrow$5$d$ gap of $1.7$~eV. Based on our work we find that the bottom of the conduction band in YbN is significantly disordered, manifesting in the transport and optical measurements.

YbN films were grown in a Riber Compact 21 CLS MBE system with a deposition chamber base pressure of $5\times10^{-10}$~mbar. The substrates were outgassed at $400$~$^{\text{o}}$C for at least four hours prior to their introduction to the growth chamber. The samples were grown at ambient temperature without actively heating or cooling the substrate. Unlike for the formation of many of the other \textit{Ln}Ns, Yb does not crack N\textsubscript{2} in ambient conditions~\cite{chan2020facile}. This required the YbN films to be grown in the presence of $120$~eV nitrogen ions provided by a Kaufman \& Robinson EH200 ion source. The films were capped with a passivating layer of $\approx50$~nm AlN. We used double side polished Al\textsubscript{2}O\textsubscript{3}(0001) and Si(001) substrates. For both depositions, the samples were continuously rotated at $5$ rotations per minute. The YbN and AlN deposition rates were $\approx80$ and $\approx170$~nm per hour, respectively.

\begin{figure}
    \centering
    \includegraphics[width=\linewidth]{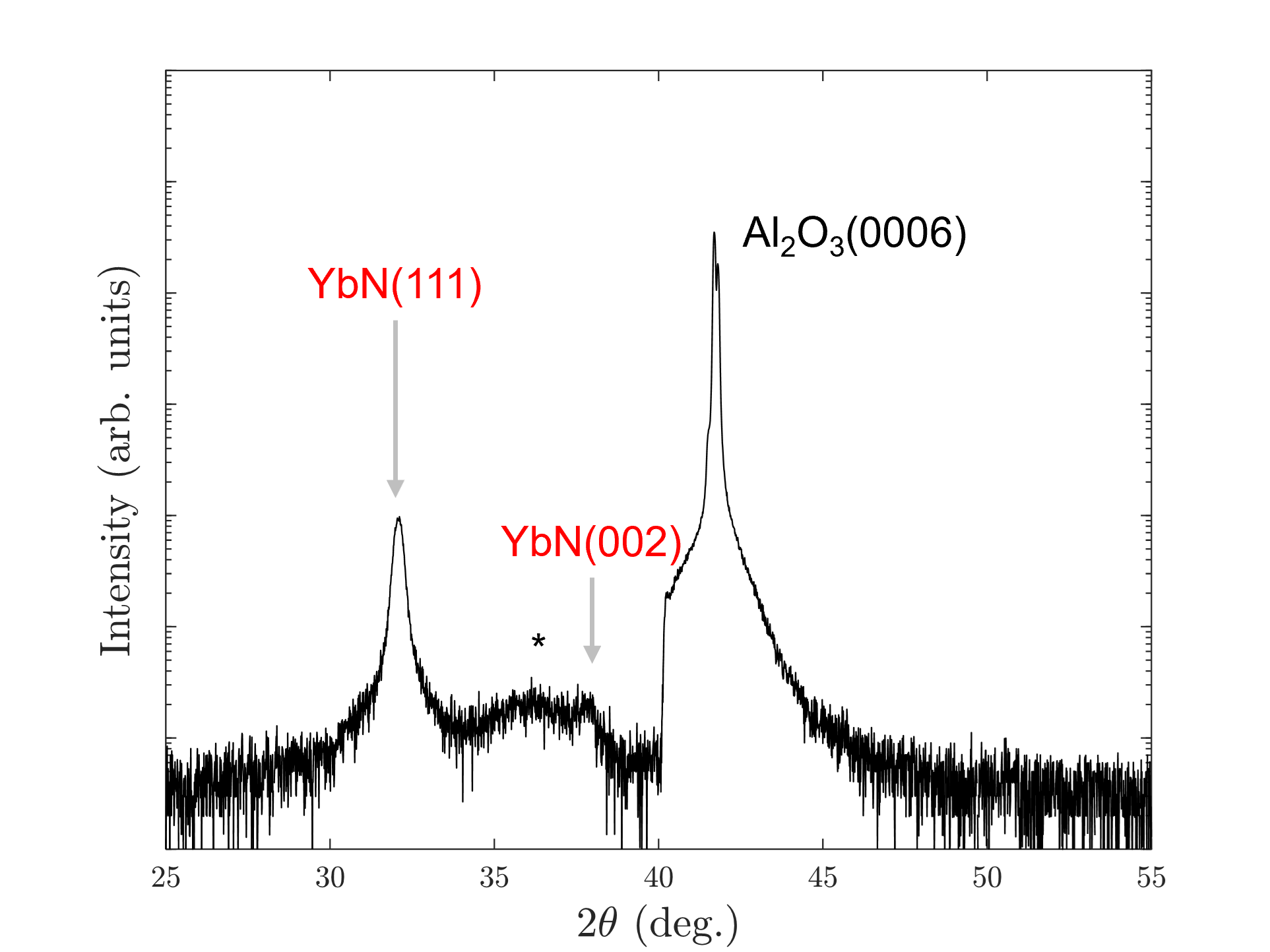}
    \caption{Angle-symmetric XRD pattern of a polycrystalline YbN film showing YbN(111) and YbN(002) diffraction planes deposited on a c-plane Al\textsubscript{2}O\textsubscript{3} substrate with a broad AlN capping layer peak marked with a *. Note the logarithmic y-scale.}
    \label{fig:YbN_XRD}
\end{figure}

The YbN film thicknesses were estimated using a Dektak profilometer by measuring the thickness of the full stack (YbN+AlN) and subtracting the thickness of the AlN which was independently determined. The crystal morphology of the YbN films were studied using x-ray diffraction (XRD) with a Rigaku Smart Lab x-ray diffractometer with a Cu x-ray source. A typical XRD measurement is shown in Figure~\ref{fig:YbN_XRD}, from which we identify that the YbN film is polycrystalline with a strong (111) texture and a lattice constant of $(4.832\pm1)$~{\AA}, when deposited on an Al\textsubscript{2}O\textsubscript{3} substrate. The minimal crystallite size derived from the width of a two-wavelength pseudo-Voigt fit to the YbN(111) peak was $(58\pm1)$~nm. The measured lattice constant is consistent with that of conducting sputter-deposited YbN rather than conducting MBE-grown YbN ~\cite{warring2014ybn,loyal2023coexistence}. We note that quite a significant variation of the lattice parameter has been observed in various \textit{Ln}Ns, appearing to depend very sensitively on the preparation conditions. While not fully understood, the larger lattice parameters occurs mainly in films produced in nitrogen rich conditions, while the smaller values (which largely align with the historical literature) are found in films which likely contain a significant concentration of nitrogen vacancies~\cite{holmes2025rare}.

\begin{figure}
    \centering
    \includegraphics[width=\linewidth]{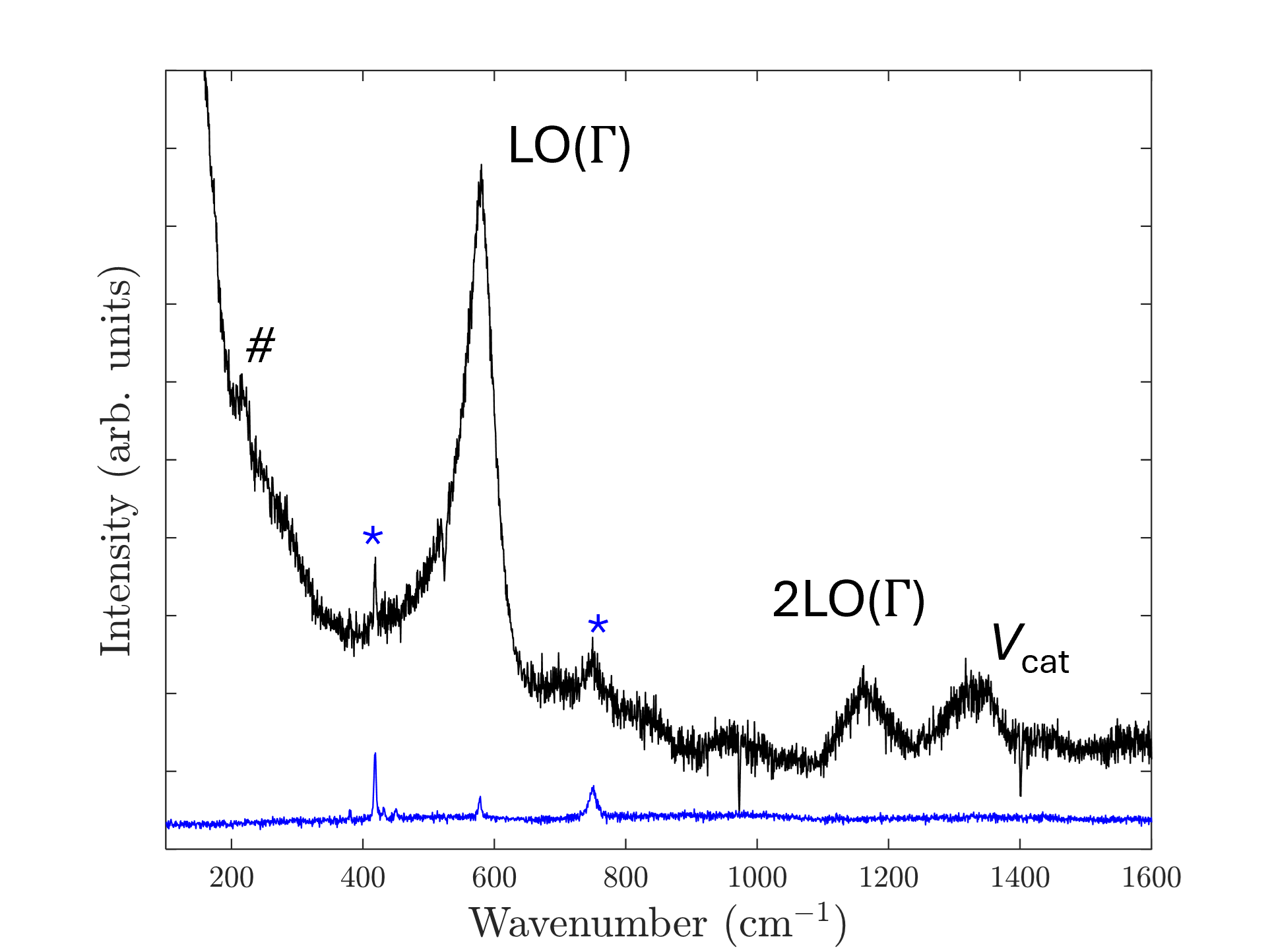}
    \caption{Raman spectra of a YbN film deposited on a Al\textsubscript{2}O\textsubscript{3} substrate (black) and a reference Al\textsubscript{2}O\textsubscript{3} measurement (blue). The most prominent feature in the spectrum is a Fröhlich polaron-enabled LO($\Gamma$) signal, seen also at double the frequency and labeled as 2LO($\Gamma$). The cation vacancy signal is marked with \textit{V}$_{\text{cat}}$.}
    \label{fig:YbN_Raman}
\end{figure}

The vibrational properties were studied using Raman spectroscopy. A Jobin-Yvon LabRam spectrometer was employed for the Raman study using a $514.5$~nm ($2.41$~eV) excitation laser. The Raman spectrum from a YbN thin film on a Al\textsubscript{2}O\textsubscript{3} substrate is shown in Figure~\ref{fig:YbN_Raman} (black), below which is a Raman spectrum that shows phonon modes originating from the substrate (blue) from~\cite{VanKoughnet2023}. The Fröhlich polaron-enabled LO($\Gamma$) phonon mode appears at $580$~cm\textsuperscript{-1}. Previous Raman measurements found signals at Raman shifts at $365$ and $575$~cm\textsuperscript{-1}~\cite{degiorgi1990electronic}. It is likely that these signals originate from Yb\textsubscript{2}O\textsubscript{3} and YbN, respectively~\cite{pandey2013anharmonic,merlin1978multiphonon}. We also find and record the presence of the corresponding second harmonic 2LO($\Gamma$) mode at $1160$~cm\textsuperscript{-1}, as well as a Yb cation vacancy-derived mode (\textit{V}\textsubscript{Yb}) at $1347$~cm\textsuperscript{-1}~\cite{markwitz2025raman}. The latter of these is a highly polarizable Raman mode has been demonstrated in other \textit{Ln}Ns as a symmetric breathing mode of six nitrogen atoms neighboring the cation vacancy defect site (previous reports did not deal with YbN). The signal marked with a hash also appears in ion-grown LuN and GdN films, suggesting an as-of-yet unidentified defect-originating mode~\cite{VanKoughnet2023}.

\begin{figure}
    \centering
    \includegraphics[width=\linewidth]{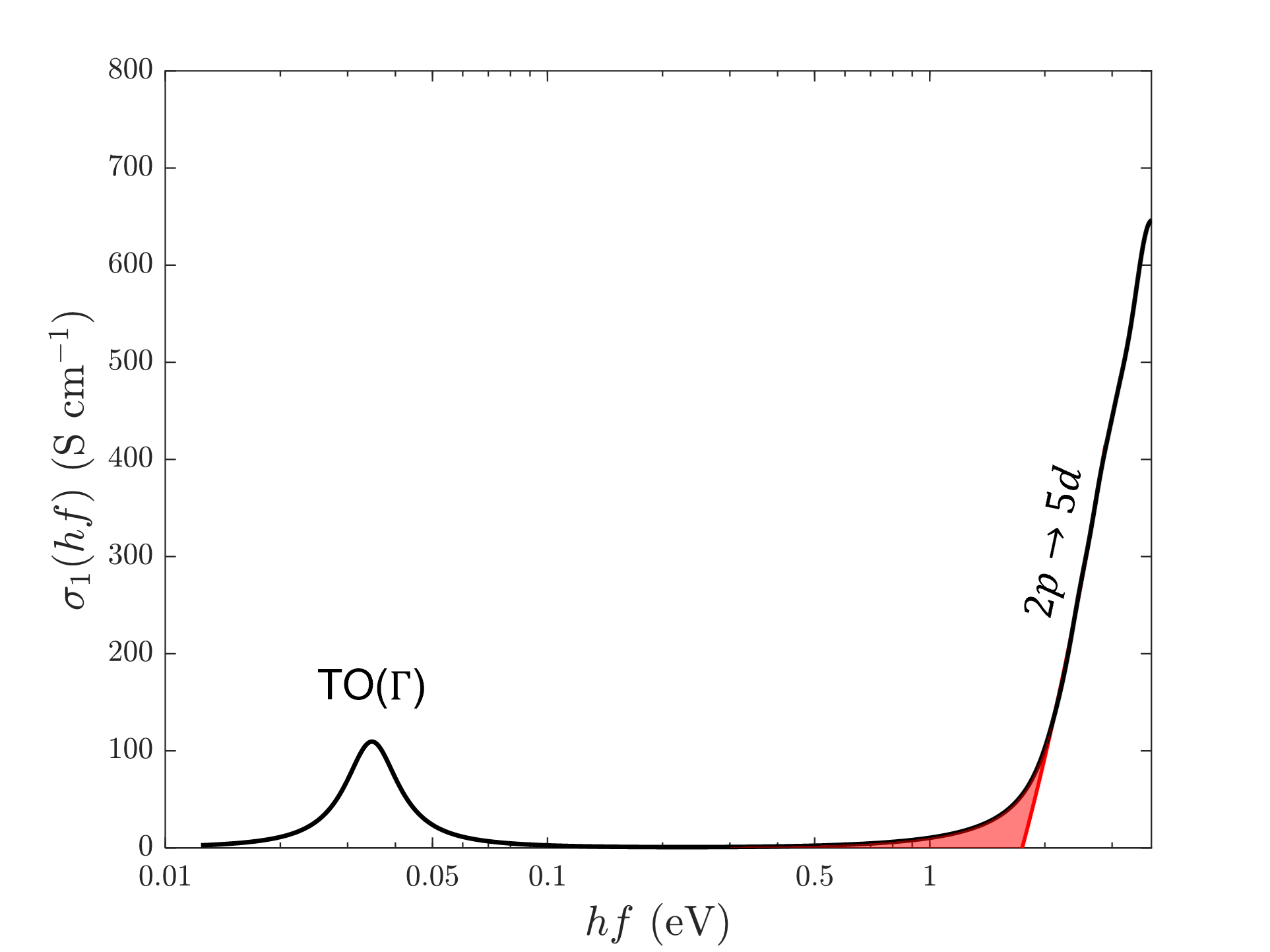}
    \caption{Real component of the optical conductivity as a function of energy for YbN. An infrared phonon absorption associated with the YbN TO($\Gamma$) mode is present at $34$~meV ($274$~cm\textsuperscript{-1}). The interband $2p\rightarrow5d$ absorption has an onset at $1.7$~eV, with an absorption tail that extends down toward $0.5$~eV highlighted in red. Note the logarithmic x-scale.}
    \label{fig:opticalConductivity}
\end{figure}

To study the optical conductivity spectrum, transmittance and reflectance measurements were conducted between $0.012$ and $3.7$~eV ($100$ and $30000$~cm\textsuperscript{-1}) using a Bruker Vertex 80v Fourier transform spectrometer. As previously described by Holmes-Hewett~\textit{et~al.}~\cite{holmes2019optical}, the software package \texttt{RefFIT} was used to generate a Kramers-Kronig consistent model of the optical conductivity function of YbN. This used the reflection and transmission obtained for an AlN film and AlN/YbN film bilayer on both Al\textsubscript{2}O\textsubscript{3} and Si substrates. We study the optical properties of YbN by measuring the frequency-dependent reflection ($R$) and transmission ($T$) of a clean substrate (both Al\textsubscript{2}O\textsubscript{3} and Si), an AlN reference film, and the complete YbN/AlN film stack. The extracted optical conductivity of the YbN film as a function of energy is depicted in Figure~\ref{fig:opticalConductivity}, showing characteristic features associated with the YbN TO($\Gamma$) mode at $34$~meV ($274$~cm\textsuperscript{-1}), in good agreement with a previously reported value~\cite{degiorgi1990electronic}. We note that the real part of the conductivity does not exhibit a free carrier term contribution at low energies due to the low conductivity of the present sample. The optical conductivity shows inter-band transitions above $1.7$~eV, which matches the direct 2$p\rightarrow$5$d$ gap of YbN~\cite{chen2025epitaxial,warring2014ybn,degiorgi1990electronic,loyal2023coexistence}. This increase in optical conductivity associated with 2$p\rightarrow$5$d$ transitions is similar to what is observed for other \textit{Ln}Ns~\cite{holmes2024spin,devese2022probing,holmes20194}. As noted in other \textit{Ln}Ns, there is a significant absorption tail below the expected optical band gap. In GdN this was understood as the effect of unintended nitrogen vacancies raising the Fermi energy into the conduction band, along with a band gap renormalization~\cite{holmes2024spin}. Here we expect the same explanation is likely for YbN.

\begin{figure}
    \centering
    \includegraphics[width=\linewidth]{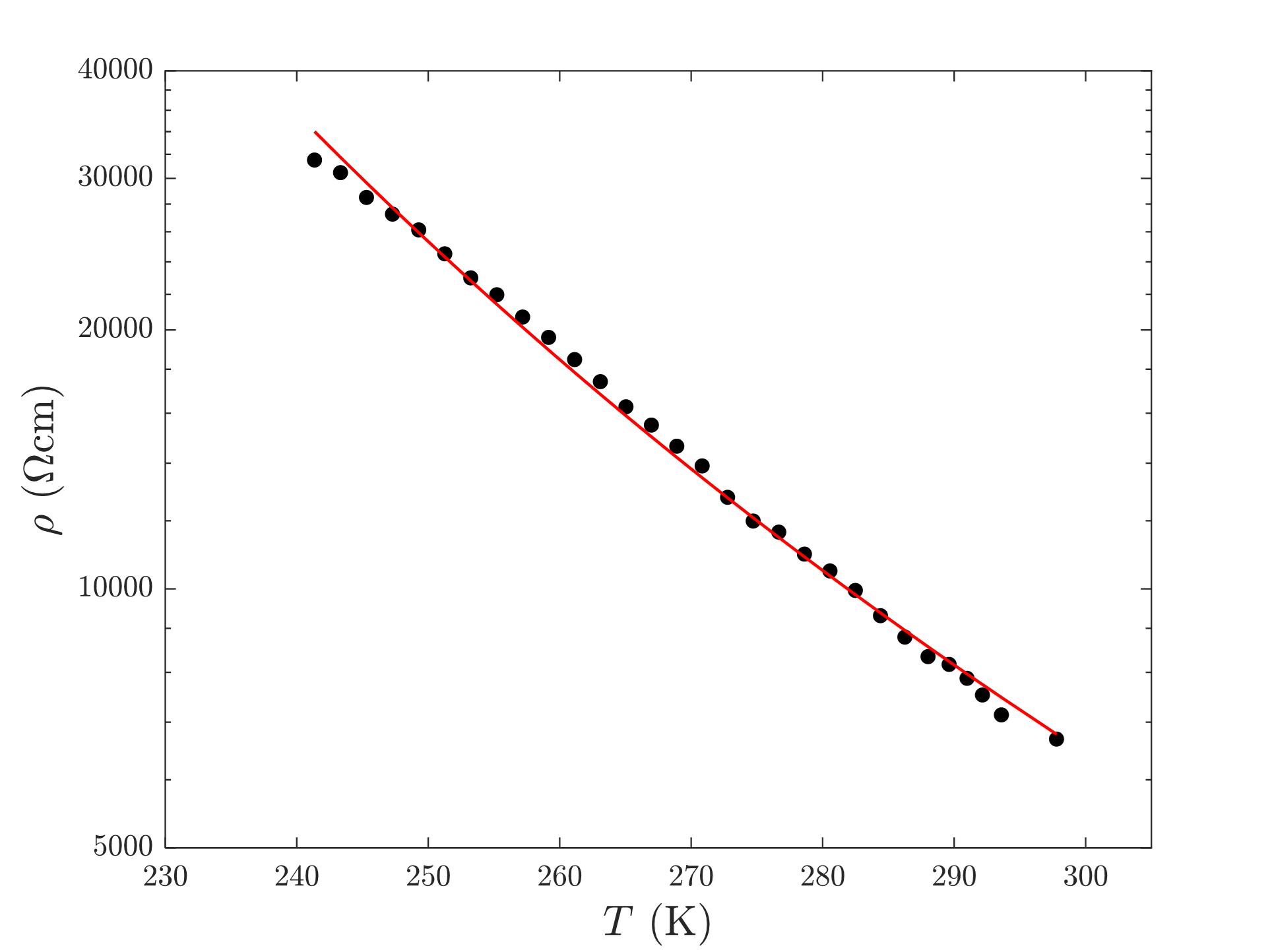}
    \caption{Temperature dependence of the resistivity (black points) between $240$ and $300$~K. The red line is an Arrhenius resistivity model described in the main text. Note the logarithmic y-scale.}
    \label{fig:YbN_RT}
\end{figure}

To further probe the electronic properties of the films the temperature dependence of the DC resistivity was studied using the conventional van der Pauw method on a square $10~\times~10$~mm\textsuperscript{2} sample of Al\textsubscript{2}O\textsubscript{3} substrate with pre-deposited contacts. The resistivity of the sample is shown in Figure~\ref{fig:YbN_RT}, showing an exponential temperature-dependence. Such a resistivity is consistent with an activation energy-based model $\rho(T)=\rho_{0}\exp{(E_{A}/(k_{B}T))}$, fitting the data with an activation energy of $(177\pm4)$~meV. This activation energy could be corresponding to transport via localized states through a hopping mechanism, or by thermal activation of carriers from localized donor states~\cite{holmes2024spin}. In the context of the optical data presented, transport via localized states is consistent with the Fermi energy residing below a mobility edge in the conduction band minimum.

Together, these results point towards YbN possessing an insulating ground state with an LO-TO splitting in-line with other \textit{Ln}Ns. Its electronic structure, however, has only been studied in the ferromagnetic state, omitting its true antiferromagnetic ground state. Experiments addressing the electronic structure of YbN however have been necessarily conducted on antiferromagnetic samples. To connect the experiment with theory this, density functional theory calculations were conducted using~\texttt{Quantum ESPRESSO}~$7.3$ on the antiferromagnetic state of YbN~\cite{giannozzi2009quantum,giannozzi2017advanced,giannozzi2020quantum}. The pseudopotentials developed by Dal~Corso~\cite{dal2014pseudopotentials} for N with 2$s$ and 2$p$ states and Topsakal and Wentzcovitch~\cite{topsakal2014accurate} for Yb with 4$f$, 5$s$, 5$p$, 5$d$, and 6$s$ states in the valence shell were used for this work. Spin-polarized calculations are essential for describing the 4$f^{13}$ occupation for Yb\textsuperscript{3+}. The calculation was found to be well-converged using a kinetic energy cutoff of $60$~Ry ($816$~eV) in conjunction with a $6\times6\times3$ $k$-point grid for the $16$ atom $1\times1\times2$ supercell AFM-III extension of the cubic unit cell of the Fm$\overline{\text{3}}$m structure. 

Rotationally symmetric Hubbard $U$ parameters were applied to the Yb 4$f$ and 5$d$ states in order to improve their description over the standard generalized-gradient-approximation~\cite{larson2007electronic,perdew1996generalized}. If not considered, the electronic structures for most all \textit{Ln}Ns are predicted to be metals with 4$f$ states partially occupied at the Fermi energy and/or with N~2$p$-\textit{Ln}~5$d$ valence-conduction band overlap. We identify the appropriate Hubbard $U$ parameter for the 4$f$ states by following the scheme of Larson~\textit{et~al.}~\cite{larson2007electronic}, and for the 5$d$ states by tuning the band gap to the strong optical absorption onset energy of $1.7$~eV at the $X$ point in the primitive Brillouin zone, taken to be associated with the onset of 2$p$-5$d$ optical transitions. The resulting Hubbard parameters are $U_{f}=8.1$~eV and $U_{d}=6.3$~eV, which are similar to the parameters which Larson~\textit{et~al.}~\cite{larson2007electronic} used for their calculations. 

\begin{figure}
    \centering
    \includegraphics[width=\columnwidth]{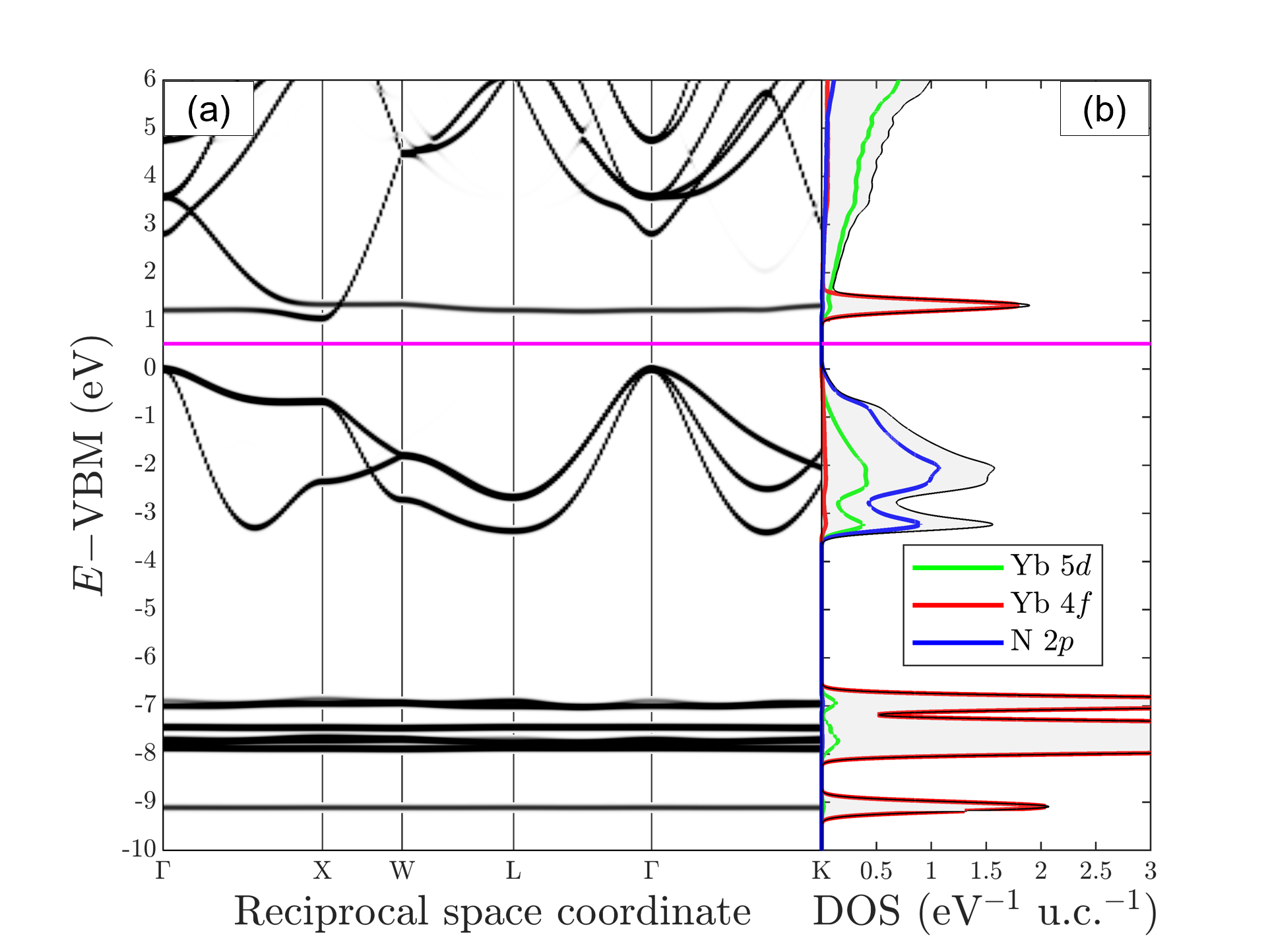}
    \caption{DFT+$U_{d}$+$U_{f}$ AFM-III YbN (a) band structure and (b) orbital-projected density of states. The Fermi energy is marked with the magenta line. The energy is referenced to the majority spin valence band maximum.}
    \label{fig:YbN_bandStructure}
\end{figure}

The corresponding band structure and density of states are shown in Figure~\ref{fig:YbN_bandStructure}(a) and Figure~\ref{fig:YbN_bandStructure}(b), respectively, where the thirteen occupied 4$f$ states lay $7-9$~eV below, and the fourteenth 4$f$ state appears $1$~eV above the valence band maximum. The computation is in qualitative agreement with the electronic structures suggested from previous measurements and calculations~\cite{degiorgi1990electronic,larson2007electronic}. Our computation suggests that there is a narrow 2$p\rightarrow$5$d$ indirect band gap of $0.8$~eV between $\Gamma$ and $X$. Additionally, the narrowest direct gap is 2$p\rightarrow$4$f$ at $\Gamma$ with a value of $1.3$~eV. The insulating ground state is in agreement with recent measurements and calculations~\cite{warring2014ybn,topsakal2014accurate,larson2007electronic}.

The energy difference in the minimum energy of the unoccupied 4$f$ and 5$d$ states is rather small, approximately $0.2$~eV. Doping YbN with nitrogen vacancies will lift the Fermi energy into the conduction band, bringing it close to the unoccupied 4$f$ states, which themselves may contribute to measured transport properties. Interestingly, there is 5$d$-4$f$ hybridization along the $\Gamma-X$ path, but a lack thereof for the $X-W$ path. This may cause the Yb\textsuperscript{3+} states to transition to Yb\textsuperscript{2+} for highly doped samples, therefore providing a mixed Yb\textsuperscript{2+/3+} valence. This can be studied by measurement of an enhanced effective mass or direct observation of Yb\textsuperscript{2+} using XPS~\cite{degiorgi1990electronic}. The originally-reported properties of what is now understood to be nonstoichiometric N-poor YbN point towards a carrier effective mass much greater than that of ScN, suggesting YbN possesses heavy Fermion or Kondo insulator physics which may be accessible via tuning the carrier concentration through control of preparation conditions~\cite{ott1985low,degiorgi1990electronic}. Previous XPS measurements on the 4$d$ states have found mixed evidence of Yb\textsuperscript{2+}~\cite{loyal2023coexistence,chen2025epitaxial}, and measurements accessing the 3$d$ core level clearly identified the presence of Yb\textsuperscript{2+}\cite{greber1987yb}. This discrepancy is likely caused by a combination of surface effects and peak separation due to the difference in binding energies of those core levels, and merits further investigation.

Chen~\textit{et~al.}~\cite{chen2025epitaxial} suggest a minimum indirect band gap of $1.64$~eV from XPS measurements by assuming that the Fermi energy resides in the conduction band and extrapolating the drop-off of the density of states at the top of the valence band. It is known from calculations on \textit{Ln}\textsuperscript{3+}N\textsuperscript{3-} species that the valence band energy at $\Gamma$ is raised relative to the $X$ point for systems with occupied 4$f$ T\textsubscript{2u} states, as was calculated for example in NdN~\cite{larson2007electronic}. It remains possible that the extrapolated VBM that Chen~\textit{et~al.}~\cite{chen2025epitaxial} identify is the sharp drop-off of the N 2$p$ density of states for energies above the $X$-point, and the true indirect band gap is smaller than they estimate. This is because the density of states reduces significantly at valence band states toward the $\Gamma$-point as we find in our calculation. Their result for the indirect band gap ($1.64$~eV) is greater than our calculated $\Gamma$-$X$ band gap of $0.8$~eV. 

When attempting to identify the presence of Yb\textsuperscript{2+} in YbN, it is worth drawing parallels with EuN for which the compound exhibits a multiple valence (Eu\textsuperscript{2+/3+}) depending on the sample preparation conditions conditions~\cite{richter2011electronic,ruck2011magnetic,le2013europium}. This is because as the EuN is doped with nitrogen vacancies, Eu\textsuperscript{3+} (4$f$\textsuperscript{6}) converts to Eu\textsuperscript{2+} (4$f$\textsuperscript{7}) with a distinct difference in magnetic signature converting from an $L=3$, $S=3$ and $J=0$ state to a $L=0$, $S=7/2$ and $J=7/2$ state by Hund's rules. It is possible that something similar may occur with non-stoichiometric YbN where Yb\textsuperscript{3+} (4$f$\textsuperscript{13}) may transition to Yb\textsuperscript{2+} (4$f$\textsuperscript{14}) configuration, i.e., from $L=3$, $S=1/2$ and $J=7/2$ state to a $L=0$, $S=0$ and $J=0$ nonmagnetic ground state by Hund's rules. There is also direct experimental evidence that the Kondo effect, i.e., heavy fermion transport, takes place in EuN~\cite{le2013europium}. Yb-based multivalent compounds such as the prototypical YbAl\textsubscript{3} can exhibit highly non-standard Fermi surfaces due to valence fluctuations, giving rise to heavy fermion transport properties~\cite{chatterjee2017lifshitz,blyth1993temperature}. Doping YbN with electrons may be a method to access those heavy fermion transport properties originating from a multivalent Yb\textsuperscript{2+/3+} state.


In summary, the electronic structure of antiferromagnetic YbN remains under debate. We elucidate a few points by a report on some properties of low-conductivity YbN deposited using ion-assisted molecular beam epitaxy which exhibited an exponential temperature-dependence of resistivity, likely associated with hopping transport of in-gap defect states, which is in general agreement with our optical data which suggests a disordered conduction band minimum. Derived from optical measurements on low conductivity YbN the direct gap at the $X$ point is measured to be $1.7$~eV, with the DFT computation suggesting an indirect gap of $0.8$~eV. The sample shows distinct TO($\Gamma$) and LO($\Gamma$) mode frequencies of 282~cm\textsuperscript{-1} and 580~cm\textsuperscript{-1}, in addition to a \textit{V}\textsubscript{Yb}-derived Raman signal at 1347~cm\textsuperscript{-1}.

It would be useful to harness control over the carrier concentration to access the mixed-valence Yb\textsuperscript{2+/3+} system and to explore the physics of competing exchange energies. This could be studied by alloying YbN with a metal donor or by doping YbN with another \textit{Ln} that may provide strong ferromagnetic spin coupling to YbN. The resulting exotic transport properties brought upon by the competing energy scales can be used as a platform to further fundamental research into the properties of these states of matter.

\section*{Acknowledgments}

The authors have benefitted from many informative discussions with E.~X.~M.~Trewick, B.~J.~Ruck, and H.~J.~Trodahl. This research was supported by the New Zealand Ministry of Business, Innovation and Employment (Grant No.~RSCHTRUSTVIC2447). The MacDiarmid Institute is supported under the New Zealand Centres of Research Excellence programme. W.~F.~Holmes-Hewett is supported by the Royal Society under the New Zealand Mana T\={u}\={a}papa Future Leader Fellowship (Contract No.~MTP-VUW2402). The computations were performed on R$\overline{\text{a}}$poi, the high performance computing facility of Victoria University of Wellington.

\bibliography{Ram}

\end{document}